# Optimal sequential transmission over broadcast channel with nested feedback

Aditya Mahajan
Department of Electrical Engineering
Yale University, New Haven, CT.
aditya.mahajan@yale.edu date

*Abstract*— We consider the optimal design of sequential transmission over broadcast channel with nested feedback. Nested feedback means that the channel output of the outer channel is also available at the decoder of the inner channel. We model the communication system as a decentralized team with three decision makers—the encoder and the two decoders. Structure of encoding and decoding strategies that minimize a total distortion measure over a finite horizon are determined. The results are applicable for real-time communication as well as for the information theoretic setup.

## I. Problem formulation and main result

In this paper, we study real-time broadcast of correlated sources over physically degraded channel with nested noiseless feedback. The communication system is shown in Figure 1. It operates in discrete time for a horizon $T$.

The source is a first-order time-homogeneous Markov chain. The source outputs $(U_t, V_t)$ take values in $\mathcal{U} \times \mathcal{V}$. The initial output of the source is distributed according to $P_{U_1 V_1}$; the transition matrix of the source is $P_{UV}$.

The source output is transmitted over a discrete memoryless broadcast channel that is physically degraded. Let $X_t \in \mathcal{X}$ denote the channel input at time $t$ and $(Y_t, Z_t) \in \mathcal{Y} \times \mathcal{Z}$ denote the channel outputs at time $t$. Since the channel is memoryless, we have

$$\Pr(Y_t = y_t, Z_t = z_t \mid U^t = u^t, V^t = v^t,$$
$$X^t = x^t, Y^{t-1} = y^{t-1}, Z^{t-1} = z^{t-1})$$
$$= \Pr(Y_t = y_t, Z_t = z_t \mid X_t = x_t)$$
$$=: Q_{YZ|X}(y_t, z_t | x_t).$$

Moreover, the channel is physically degraded, so

$$Q_{YZ|X}(y, z|x) = Q_{Y|X}(y|x) Q_{Z|X}(z|x).$$

Sometimes it is more convenient to describe the channel in a functional form as

$$Y_t = q_1(X_t, N_{1,t}), \quad Z_t = q_2(Y_t, N_{2,t}).$$

The channel noises $\{N_{1,t}, t = 1, \ldots, T\}$ and $\{N_{2,t}, t = 1, \ldots, T\}$ are i.i.d. sequences that are mutually independent and also independent of the source outputs. The channel functions $q_1$ and $q_2$ and the distribution of the noises are consistent with the conditional distributions $Q_{Y|X}$ and $Q_{Z|Y}$.

The communication system consists of an encoder and two decoders, all of which operate causally and in real-time. The decoder that receives $Y_t$ is called the *inner decoder* while the decoder that receives $Z_t$ is called the *outer decoder*. The channel is used with feedback, i.e., $Y_t$ is available to the encoder after a unit delay and $Z_t$ is available to the encoder and the inner decoder after a unit delay.

The encoder is described by an encoding strategy $c^T :=$ $(c_1, \ldots, c_T)$ where

$$c_t : \mathcal{U}^t \times \mathcal{V}^t \times \mathcal{X}^{t-1} \times \mathcal{Y}^{t-1} \times \mathcal{Z}^{t-1} \mapsto \mathcal{X}.$$

The encoded symbol at time $t$ is generated according to the encoding rule $c_t$ as follows

$$X_t = c_t(U^t, V^t, X^{t-1}, Y^{t-1}, Z^{t-1}). \tag{1}$$

The inner decoder is described by a decoding strategy $g_1^T :=$ $(g_{1,1}, \ldots, g_{1,T})$ where

$$g_{1,t} : \mathcal{Y}^t \times \mathcal{Z}^{t-1} \mapsto \hat{\mathcal{U}}.$$

Similarly, the outer decoder is described by a decoding strategy $g_2^T :=$ $(g_{2,1}, \ldots, g_{2,T})$ where

$$g_{2,t} : \mathcal{Z}^t \mapsto \hat{\mathcal{V}}.$$

Thus, the decoded symbols at time $t$ are generated as follows

$$\hat{U}_t = g_{1,t}(Y^t, Z^{t-1}); \tag{2}$$
$$\hat{V}_t = g_{2,t}(Z^t). \tag{3}$$

The fidelity of reconstruction at the two decoders is determined by distortion functions $\rho_{1,t} : \mathcal{U} \times \hat{\mathcal{U}} \mapsto [0, \rho_{\max}]$ and $\rho_{2,t} : \mathcal{V} \times \hat{\mathcal{V}} \mapsto [0, \rho_{\max}]$, where $\rho_{\max} < \infty$. For any communication strategy $(c^T, g_1^T, g_2^T)$, the system incurs an expected distortion given by

$$J(c^T, g_1^T, g_2^T)$$
$$:= \mathbb{E}^{(c^T, g_1^T, g_2^T)} \left\{ \sum_{t=1}^T \left[ \rho_{1,t}(U_t, \hat{U}_t) + \rho_{2,t}(V_t, \hat{V}_t) \right] \right\}. \tag{4}$$

We are interested in the optimal design of the above communication system. Specifically, we are interested in the following optimization problem.

*Problem 1:* Given the statistics of the source and the channel, the distortion functions $\rho_{1,t}$ and $\rho_{2,t}$, and the time horizon $T$, choose a communication strategy $(c^{*T}, g_1^{*T}, g_2^{*T})$, with encoders of the form (1) and decoders of the form (2) and (3), such that $(c^{*T}, g_1^{*T}, g_2^{*T})$ minimizes the expected total distortion given by (4).

Since the alphabets $\mathcal{U}, \mathcal{V}, \mathcal{X}, \mathcal{Y},$ and $\mathcal{Z}$ are finite, the number of communication strategies are finite. Therefore, in principle, we can evaluate the performance of all of them and choose the one with the best performance. Consequently, Problem 1 is well posed.

The domain of the encoding and decoding functions of the form (1), (2), (3) increases exponentially with time. As a result, the number of communication strategies increase doubly exponentially with time. Furthermore, implementing a communication strategy for a large horizon becomes difficult. In this paper, we find structural properties of optimal communication strategies that will allow us to "compress" the information available at a node to a sufficient statistic. The size of these sufficient statistics does not increase with time; therefore, the domain of the encoding and decoding functions does not change with time. Consequently, implementing a communication strategy that is such a form is easier.

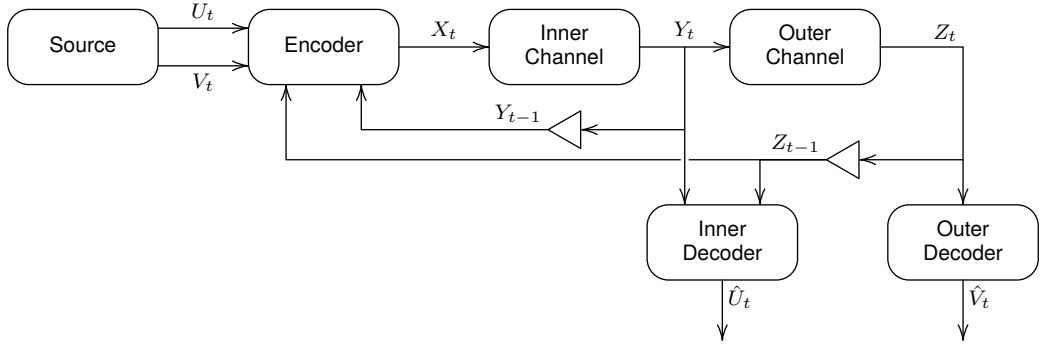

Fig. 1. A broadcast communication system with feedback

A simplified version of the structural results is stated below. The more formal version of the structural result and along with its derivation is presented in Section II.

*Theorem 1:* Without loss of optimality, we can restrict attention to communication strategies where the encoding rule is of the form

$$c_t : \mathcal{U} \times \mathcal{V} \times \Delta(\mathcal{U} \times \mathcal{V}) \times \Delta\big(\Delta(\mathcal{U} \times \mathcal{V})\big) \mapsto \mathcal{X}, \quad (5)$$

and the decoding functions are of the form

$$g_{1,t} : \Delta(\mathcal{U}) \mapsto \hat{\mathcal{U}}, \quad (6)$$

$$g_{2,t} : \Delta(\mathcal{V}) \mapsto \hat{\mathcal{V}}. \quad (7)$$

Specifically, a strategy of the following form is optimal.

$$X_t = c_t\Big(U_t, V_t, \Pr\Big(U_{t-1}, V_{t-1} \,\Big|\, Y^{t-1}, Z^{t-1}\Big),$$
$$\Pr\Big(U_{t-1}, V_{t-1}, \Pr\Big(U_{t-1}, V_{t-1} \,\Big|\, Y^t, Z^{t-1}\Big) \,\Big|\, Z^{t-1}\Big)\Big),$$
$$\hat{U}_t = g_{1,t}\Big(\Pr\Big(U_t \,\Big|\, Y^t, Z^{t-1}\Big)\Big),$$
$$\hat{V}_t = g_{2,t}\Big(\Pr\Big(V_t \,\Big|\, Z^t\Big)\Big).$$

These structural results imply that we can restrict attention to communication strategies where the domain of the encoding and the decoding functions is not increasing with time. This restriction allows us to write the communication strategy in a recursive form. Due to lack of space, we cannot present the recursive form in this paper.

Consider the following special case.

- $P_{U_1 V_1}$ is a uniform distribution over $\mathcal{U} \times \mathcal{V}$. $P_{UV}$ is an identity matrix and $\rho_{i,t} \equiv 0$ for $t = 0, \ldots, T-1$ and

$$\rho_{i,T}(w, \hat{w}) = \begin{cases} 0, & \text{if } w \neq \hat{w}; \\ 1, & \text{otherwise.} \end{cases}$$

In this case the source does not change with time. So, we drop the subscripts and denote the source output by $U$ and $V$. The total cost (4) of a communication strategy $(c^T, g_1^T, g_2^T)$ equals

$$J(c^T, g_1^T, g_2^T) = \Pr(U \neq U_T) + \Pr(V \neq V_T) \quad (8)$$

This special case is (almost[1]) identical to the information theoretic setup of communicating over broadcast channels [1]. Therefore, the structural results presented in this paper are also applicable to the information theoretic setup.

The capacity of degraded broadcast channel was computed in [2, 3]. For degraded broadcast channels, feedback does not increase capacity [4, 5]. Nonetheless, as in point-to-point communication, feedback can simplify the communication scheme. We believe that the structural results presented in this paper will be useful for finding recursive schemes that can achieve capacity of broadcast channels with feedback.

## II. STRUCTURAL RESULTS

The domain of the encoding functions of the form (1) increases with time because of three elements: the source outputs $(U^t, V^t)$, the channel inputs $X^{t-1}$, and the channel outputs $(Y^{t-1}, Z^{t-1})$. The channel outputs also increase the domain of the decoding rules with time. We compress each of these elements one by one by proceedings as follows.

1. *Ignoring past source outputs and channel inputs.*
   First, we show that the past source outputs and the past channel inputs can be ignored at the encoder. Thus, without loss of optimality, we can restrict attention to encoding rules of the form
   $$c_t : \mathcal{U} \times \mathcal{V} \times \mathcal{Y}^{t-1} \times \mathcal{Z}^{t-1} \mapsto \mathcal{X}.$$
   Specifically,
   $$X_t = c_t(U_t, V_t, Y^{t-1}, Z^{t-1}).$$

2. *Compressing $Y^{t-1}$ to a sufficient statistic.*
   Next, we consider an equivalent reformulation of the problem where a coordinator chooses the encoding and the inner decoding functions. This coordinator can compress the outputs $Y^{t-1}$ of the inner channel to a sufficient statistic such that we can restrict attention to encoding and inner decoding functions of the form
   $$c_t : \mathcal{U} \times \mathcal{V} \times \Delta(\mathcal{U} \times \mathcal{V}) \times \mathcal{Z}^{t-1} \mapsto \mathcal{X},$$
   $$g_{1,t} : \mathcal{Y} \times \Delta(\mathcal{U} \times \mathcal{V}) \times \mathcal{Z}^{t-1} \mapsto \hat{\mathcal{U}}.$$
   Specifically,
   $$X_t = c_t\Big(U_t, V_t, \Pr\Big(U_t, V_t \,\Big|\, Y^{t-1}, Z^{t-1}\Big), Z^{t-1}\Big),$$
   $$\hat{U}_t = g_{1,t}\Big(Y_t, \Pr\Big(U_t, V_t \,\Big|\, Y^{t-1}, Z^{t-1}\Big), Z^{t-1}\Big).$$

3. *Compressing $Z^{t-1}$ to a sufficient statistic.*
   After that we consider an equivalent reformulation where a coordinator chooses the communication strategy. This coordinator can compress the outputs $Z^{t-1}$ of the outer channel to a sufficient statistic such that we can restrict attention to encoding and decoding functions of the form
   $$c_t : \mathcal{U} \times \mathcal{V} \times \Delta(\mathcal{U} \times \mathcal{V}) \times \Delta\big(\Delta(\mathcal{U} \times \mathcal{V})\big) \mapsto \mathcal{X},$$

---

[1]In the information theoretic setup, the probability of error is $\Pr(\{U \neq \hat{U}_T\}$ or $\{V \neq V_T\})$. As the two errors $\{U \neq U_T\}$ and $\{V \neq V_T\}$ are not independent, so (8) is not exactly the same as the information theoretic setup. Nevertheless, the two setups are essentially the same.

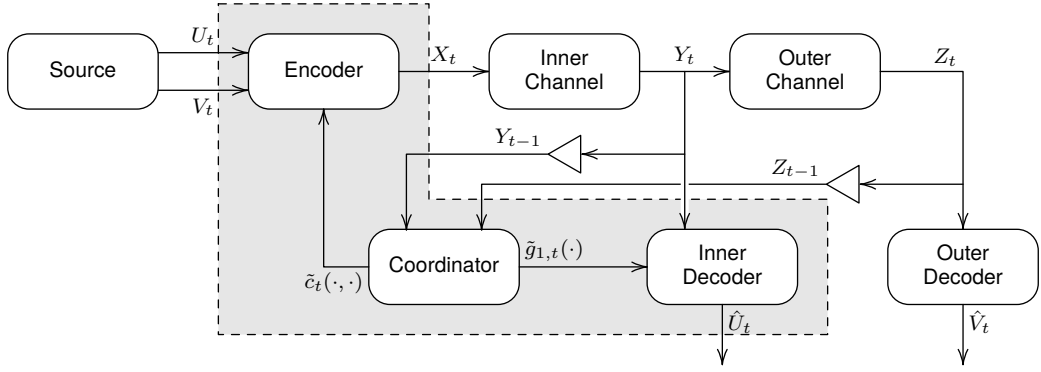

Fig. 2. An alternative formulation of the broadcast system with feedback.

$$g_{1,t}: \mathcal{Y} \times \Delta(\mathcal{U} \times \mathcal{V}) \times \Delta\big(\Delta(\mathcal{U} \times \mathcal{V})\big) \mapsto \hat{\mathcal{U}},$$
$$g_{2,t}: \Delta\big(\Delta(\mathcal{U} \times \mathcal{V})\big) \mapsto \hat{\mathcal{V}}.$$

Specifically,

$$X_t = c_t\Big(U_t, V_t, \Pr\big(U_{t-1}, V_{t-1}\,\big|\,Y^{t-1}, Z^{t-1}\big),$$
$$\Pr\Big(U_{t-1}, V_{t-1}, \Pr\big(U_{t-1}, V_{t-1}\,\big|\,Y^t, Z^{t-1}\big)\,\Big|\,Z^{t-1}\Big)\Big),$$
$$\hat{U}_t = g_{1,t}\Big(Y_t, \Pr\big(U_t, V_t\,\big|\,Y^t, Z^{t-1}\big),$$
$$\Pr\Big(U_{t-1}, V_{t-1}, \Pr\big(U_{t-1}, V_{t-1}\,\big|\,Y^t, Z^{t-1}\big)\,\Big|\,Z^{t-1}\Big)\Big),$$
$$\hat{V}_t = g_{2,t}\Big(\Pr\big(U_{t-1}, V_{t-1},$$
$$\Pr\big(U_{t-1}, V_{t-1}\,\big|\,Y^t, Z^{t-1}\big)\,|\,Z^{t-1}\big)\Big).$$

4. *A smaller sufficient statistic for the decoders.*

At this stage, we already have a structural result where the domain of the communication strategy is not increasing with time. The decoding rules can nevertheless be further simplified to

$$g_{1,t}: \Delta(\mathcal{U}) \mapsto \hat{\mathcal{U}},$$
$$g_{2,t}: \Delta(\mathcal{V}) \mapsto \hat{\mathcal{V}}.$$

Specifically,

$$\hat{U}_t = g_{1,t}(\Pr\big(U_t\,\big|\,Y^t, Z^{t-1}\big)),$$
$$\hat{V}_t = g_{2,t}(\Pr\big(V_t\,\big|\,Z^t\big)).$$

Below, we elaborate on each of these steps.

### A. Ignoring past source outputs and channel inputs

The past source outputs and the channel inputs can be ignored at the encoder. Specifically, we have the following.

*Proposition 1:* Without loss of optimality, we can restrict attention to encoding rules of the form

$$c_t: \mathcal{U} \times \mathcal{V} \times \mathcal{Y}^{t-1} \times \mathcal{Z}^{t-1} \mapsto \mathcal{X}$$

with

$$X_t = c_t(U_t, V_t, Y^{t-1}, Z^{t-1}). \tag{9}$$

*Proof.* Define $R_t = (U_t, V_t, Y^{t-1}, Z^{t-1})$. It can be verified that

$$\Pr\big(R_{t+1}\,\big|\,U^t, V^t, X^t, Y^{t-1}, Z^{t-1}\big) = \Pr\big(R_{t+1}\,|\,R_t, X_t\big).$$

Furthermore,

$$\mathbb{E}\Big\{\rho_{1,t}(U_t, \hat{U}_t) + \rho_{2,t}(V_t, \hat{V}_t)\,\Big|\,U^t, V^t, X^t, Y^{t-1}, Z^{t-1}\Big\}$$
$$= \mathbb{E}\Big\{\rho_{1,t}(U_t, \hat{U}_t) + \rho_{2,t}(V_t, \hat{V}_t)\,\Big|\,R_t, X_t\Big\}.$$

Thus, the process $\{R_t, t = 1, \ldots, T\}$ is a controlled Markov chain given $X_t$. Further, the conditional expectation of the instantaneous distortion given $(R_t, X_t)$ depends only on $(R_t, X_t)$. The state $R_t$ of the chain is perfectly observed at the encoder (which has to choose $X_t$). Hence, the results of Markov decision theory [6] imply that restricting attention to encoders of the form (9) does not incur a loss of optimality. □

From now on, we will assume that the encoder is of the form (9). Thus, we can simplify Problem 1 as follows.

*Problem 2:* Under the assumptions of Problem 1, find optimal communication strategy $(c^{*T}, g_1^{*T}, g_2^{*T})$ with encoders of the form (9) and decoders of the form (2) and (3).

### B. Compressing $Y^{t-1}$ to a sufficient statistic

To find a sufficient statistic for $Y^{t-1}$, we proceed as follows.
1. Fix a decoding policy of the outer decoder and formulate a stochastic control problem from the point of view of a coordinator that observes $(Y^{t-1}, Z^{t-1})$.
2. Show that the coordinator's problem is equivalent to the original problem. Specifically, any strategy for the coordinator's problem can be implemented in the original problem in the absence of a physical coordinator. Contrariwise, any strategy of the original problem can be implemented by the coordinator.
3. Identify a controlled Markov process that is observed at the controller and use that to identify a sufficient statistic for $Y^{t-1}$.

Below we elaborate on each of these stages.

*Stage 1*

Consider the following modified problem. In addition to the encoders and the two decoders, assume that a coordinator is present in the system that knows $(Y^{t-1}, Z^{t-1})$ at time $t$. This information $(Y^{t-1}, Z^{t-1})$ is the information shared between the encoder and the inner decoder at time $t$. Based on this shared information, the coordinator decides *partial encoding and decoding functions*

$$\tilde{c}_t : \mathcal{U} \times \mathcal{V} \mapsto \mathcal{X},$$
$$\tilde{g}_{1,t} : \mathcal{Y} \mapsto \hat{\mathcal{U}}.$$

These functions map the *private information* of the encoder and the decoder to their decisions. The coordinator then informs the encoder and the inner decoder of $\tilde{c}_t$ and $\tilde{g}_{1,t}$. The encoder and the inner decoder use their respective partial function to choose an action as follows.

$$X_t = \tilde{c}_t(U_t, V_t), \tag{10}$$

$$\hat{U}_t = \tilde{g}_{1,t}(Y_t). \tag{11}$$

The dynamics of the source and the channel along with the operation of the outer decoder are the same as in the original problem (Problem 2). At the next time step, the coordinator observes $(Y_t, Z_t)$ and selects the partial functions $(\tilde{c}_{t+1}, \tilde{g}_{1,t+1})$. The system proceeds sequentially in this manner until horizon $T$. The block diagram of the system is shown in Figure 2.

In the above formulation, there are two decision makers: the coordinator and the outer decoder. The encoder and the inner decoder simply carry out the computations prescribed in (10) and (11). At time $t$, the coordinator knows the shared information $(Y^{t-1}, Z^{t-1})$ and all the past partial functions $(\tilde{c}^{t-1}, \tilde{g}_1^{t-1})$. The coordinator's decision rule $\tilde{\phi}_t$ maps this information to its decisions, that is,

$$(\tilde{c}_t, \tilde{g}_{1,t}) = \tilde{\phi}_t(Y^{t-1}, Z^{t-1}, \tilde{c}^{t-1}, \tilde{g}_1^{t-1}). \tag{12}$$

The choice of $\tilde{\phi}^T$ is called a coordination strategy. The expected total distortion of a strategy $(\tilde{\phi}^T, g_2^T)$ is given by

$$\tilde{J}(\tilde{\phi}^T, g_2^T) = \mathbb{E}^{(\tilde{\phi}^T, g_2^T)} \left\{ \sum_{t=1}^T \left[ \rho_{1,t}(U_t, \hat{U}_t) + \rho_{2,t}(V_t, \hat{V}_t) \right] \right\}. \tag{13}$$

We are interested in the optimal design of the above system, which can be set up as the following optimization problem.

*Problem 3:* Under the assumptions of Problem 1, find a strategy $(\tilde{\phi}^{*T}, g_2^{*T})$ with $\tilde{\phi}^{*T}$ of the form (12) and $g_{2,t}$ of the form (3) such that $(\tilde{\phi}^{*T}, g_2^{*T})$ minimizes the expected total distortion given by (13).

*Stage 2*

Now we show that Problem 3 is equivalent to Problem 2. Specifically, we show that any strategy $(c^T, g_1^T, g_2^T)$ for Problem 2 can be implemented by the coordinator in Problem 3 and any strategy $(\tilde{\phi}^T, g_2^T)$ for Problem 3 can be implemented by the encoder and the decoders in Problem 2.

Any strategy $(c^T, g_1^T, g_2^T)$ in Problem 2 can be implemented in Problem 3 as follows. Keep the outer decoding strategy $g_2^T$ as is. At time $t$, the coordinator selects partial functions $(\tilde{c}_t, \tilde{g}_{1,t})$ using the shared information $(y^{t-1}, z^{t-1})$ as follows. Let

$$(\tilde{c}_t, \tilde{g}_{1,t}) = \tilde{\phi}_t(y^{t-1}, z^{t-1}) \tag{14a}$$

where $\tilde{\phi}_t$ is chosen such that

$$\tilde{c}_t(u_t, v_t) = c_t(u_t, v_t, y^{t-1}, z^{t-1}), \tag{14b}$$

$$\tilde{g}_{1,t}(y_t) = g_{1,t}(y_t, y^{t-1}, z^{t-1}). \tag{14c}$$

Now consider Problems 2 and 3. Use strategy $(c^T, g_1^T, g_2^T)$ in Problem 2 and strategy $(\tilde{\phi}^T, g_2^T)$ in Problem 3 where $\tilde{\phi}^T$ is given by (14). Consider a specific realization of the source output $\{(U_t, V_t), t = 1, \ldots, T\}$ and the channel noise $\{(N_{1,t}, N_{2,t}), t = 1, \ldots, T\}$. The choice of $\tilde{\phi}^T$ according to (14) implies that the channel inputs $\{X_t, t = 1, \ldots, T\}$, the channel outputs $\{(Y_t, Z_t), t = 1, \ldots, T\}$, and the reconstructions $\{(\hat{U}_t, \hat{V}_t), t = 1, \ldots, T\}$ are identical in Problems 2 and 3. Thus, any strategy $(c^T, g_1^T, g_2^T)$ in Problem 2 can be implemented by the coordinator in Problem 3 by using a coordination strategy given by (14). Furthermore, the total expected distortion in both cases is identical.

By a similar argument, any strategy $(\tilde{\phi}^T, g_2^T)$ for Problem 3 can be implemented in Problem 2 as follows. Keep the outer decoding strategy as is. At time $t$,

$$(\tilde{c}_t, \tilde{g}_{1,t}) = \tilde{\phi}_t(y^{t-1}, z^{t-1}, \tilde{c}^{t-1}, \tilde{g}_1^{t-1}).$$

By recursively substituting the values of $\tilde{c}^{t-1}$ and $\tilde{g}_1^{t-1}$, we can write this as

$$(\tilde{c}_t, \tilde{g}_{1,t}) = \tilde{\phi}_t(y^{t-1}, z^{t-1}, \phi_{t-1}(y^{t-2}, z^{t-2}, \ldots, \phi_1))$$
$$=: F_t(\tilde{\phi}^t, y^{t-1}, z^{t-1}). \tag{15a}$$

Let $F_{1,t}(\cdot)$ and $F_{2,t}(\cdot)$ denote the first and second components of $F_t(\cdot)$, i.e.,

$$\tilde{c}_t = F_{1,t}(\tilde{\phi}^t, y^{t-1}, z^{t-1}),$$
$$\tilde{g}_{1,t} = F_{2,t}(\tilde{\phi}^t, y^{t-1}, z^{t-1}).$$

Then, use the following encoding and inner decoding strategy in Problem 2:

$$c_t(u_t, v_t, y^{t-1}, z^{t-1}) = F_{1,t}(\tilde{\phi}^t, y^{t-1}, z^{t-1})(u_t, v_t), \tag{15b}$$

$$g_{1,t}(y^t, z^{t-1}) = F_{2,t}(\tilde{\phi}^t, y^{t-1}, z^{t-1})(y_t). \tag{15c}$$

Now consider Problems 3 and 2. Use strategy $(\tilde{\phi}^T, g_2^T)$ in Problem 3 and strategy $(c^T, g_1^T, g_2^T)$ in Problem 2 where $(c^T, g_1^T)$ is given by (15). Consider a specific realization of the source output $\{(U_t, V_t), t = 1, \ldots, T\}$ and the channel noise $\{(N_{1,t}, N_{2,t}), t = 1, \ldots, T\}$. The choice of $(c^T, g_1^T)$ according to (15) implies that the channel inputs $\{X_t, t = 1, \ldots, T\}$, the channel outputs $\{(Y_t, Z_t), t = 1, \ldots, T\}$, and the reconstructions $\{(\hat{U}_t, \hat{V}_t), t = 1, \ldots, T\}$ are identical in Problems 3 and 2. Thus, any strategy $(\tilde{\phi}^T, g_2^T)$ in Problem 3 can be implemented by the encoder and decoders in Problem 2 by using a strategy given by (15). Furthermore, the total expected distortion in both cases is identical.

The above arguments show that Problems 2 and 3 are equivalent. We now identify a sufficient statistic for compressing $Y^{t-1}$ in Problem 3.

*Stage 3*

We first define the following.

*Definition 1:* For any choice of $\tilde{c}^T$, define

$$\Xi_t(Y^t, Z^t; \tilde{c}^t) := \Pr^{\tilde{c}^t}\left(U_t, V_t \,\middle|\, Y^t, Z^t\right). \tag{16}$$

For any choice of $\tilde{c}^t$, the channel outputs $(Y^t, Z^t)$ are random variables (measurable on the probability space on which the source outputs and the channel noise are defined). Given a realization $(y^t, z^t)$ of $(Y^t, Z^t)$, the realization $\xi_t$ of $\Xi_t$ is a conditional probability on $(U_t, V_t)$ given $(y^t, z^t)$. On the other hand, when $(Y^t, Z^t)$ are random variables, $\Xi_t$ is a random variable taking values in $\Delta(\mathcal{U} \times \mathcal{V})$. Moreover, $\Xi_t$ is related to $\Xi_{t-1}$ as follows.

*Proposition 2:* Fix arbitrary partial encoding functions $\tilde{c}^T$. Then, the update of $\Xi_t$ is given by

$$\xi_t(y^t, z^t; \tilde{c}^t) = f_1\left(\xi_{t-1}(y^{t-1}, z^{t-1}; \tilde{c}^{t-1}), y_t, z_t, \tilde{c}_t\right) \tag{17}$$

where $f_1(\cdot)$ is given by

$$f_1(\xi, y, z, \hat{c})(u, v) = \sum_{(u', v') \in \mathcal{U} \times \mathcal{V}} P_{UV}(u, v | u', v')$$
$$\times \frac{Q_{Z|Y}(z|y) Q_{Y|X}(y | \tilde{c}(u', v')) \xi(u', v')}{\sum_{(y', z') \in \mathcal{Y} \times \mathcal{Z}} Q_{Z|Y}(z' | y') Q_{Y|X}(y' | \tilde{c}(u', v')) \xi(u', v')}.$$

*Proof.* This is a direct consequence of Definition 1 and Bayes's rule. □

$\Xi_{t-1}$ is a sufficient statistic for $Y^{t-1}$ in Problem 3. In particular, we have the following result.

*Proposition 3:* Arbitrarily fix the outer decoding strategy $g_2^T$. Then, in Problem 3, without loss of optimality we can restrict atten-

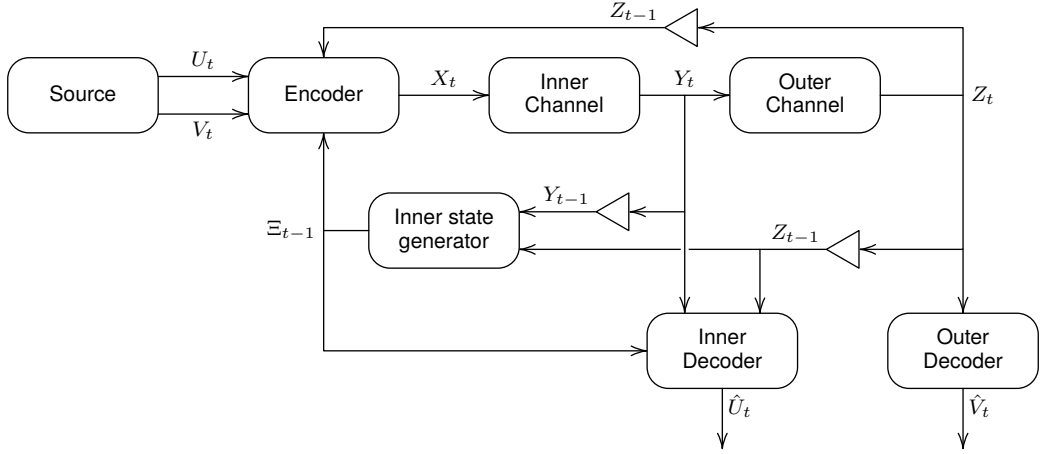

Fig. 3. The broadcast system with simplified encoder and inner decoder.

tion to a coordination strategy of the form

$$\tilde{\phi}_t : \Delta(\mathcal{U} \times \mathcal{V}) \times \mathcal{Z}^{t-1} \mapsto ((\mathcal{U} \times \mathcal{V} \mapsto \mathcal{X}), (\mathcal{Y} \mapsto \hat{\mathcal{U}}))$$

with

$$(\tilde{c}_t, \tilde{g}_{1,t}) = \tilde{\phi}_t(\Xi_{t-1}, Z^{t-1}). \tag{18}$$

Since Problems 2 and 3 are equivalent, the above implies that in Problem 2, without loss of optimality we can restrict attention to encoding and inner decoding strategies of the form

$$c_t : \mathcal{U} \times \mathcal{V} \times \Delta(\mathcal{U} \times \mathcal{V}) \times \mathcal{Z}^{t-1} \mapsto \mathcal{X},$$

$$g_{1,t} : \mathcal{Y} \times \Delta(\mathcal{U} \times \mathcal{V}) \times \mathcal{Z}^{t-1} \mapsto \hat{\mathcal{U}}.$$

with

$$X_t = c_t(U_t, V_t, \Xi_{t-1}, Z^{t-1}), \tag{19}$$

$$\hat{U}_t = g_{1,t}(Y_t, \Xi_{t-1}, Z^{t-1}). \tag{20}$$

*Proof.* Define $R_t = (\Xi_{t-1}, Z^{t-1})$. It can be verified that

$$\Pr\left(R_{t+1} \mid R^t; \tilde{c}^t\right) = \Pr\left(R_{t+1} \mid R_t; \tilde{c}_t\right).$$

Furthermore,

$$\mathbb{E}\left\{\rho_{2,t-1}(V_{t-1}, \hat{V}_{t-1}) + \rho_{1,t}(U_t, \hat{U}_t) \,\Big|\, R^t; \tilde{c}^t, \tilde{g}_1^t, g_2^t\right\}$$

$$= \mathbb{E}\left\{\rho_{2,t-1}(V_{t-1}, \hat{V}_{t-1}) + \rho_{1,t}(U_t, \hat{U}_t) \,\Big|\, R_t; \tilde{c}_t, \tilde{g}_{1,t}, g_{2,t}\right\}$$

As the outer decoder policy $g_2^T$ is fixed, the expected instantaneous cost only depends on $(R_t, \tilde{c}_t, \tilde{g}_{1,t})$. The state $R_t$ of the process is perfectly observed at the coordinator. Hence, the results of Markov decision theory [6] imply that restricting attention to coordinator strategies of the form (18) does not incur a loss of optimality. □

From now on, we will assume that the encoder and the inner decoder are of the form (19) and (20). Thus, the broadcast system can be viewed as shown in Figure 3. The system has a *inner state-generator*, which carries out the computations prescribed in (17). At time $t$, the state-generator computes $\Xi_{t-1}$ and communicates it to the encoder and the inner decoder. The encoder and the inner decoder use $\Xi_{t-1}$ along with their private information, $(U_t, V_t, Z^{t-1})$ and $(Y_t, Z^{t-1})$, respectively, to implement communication strategy of the form (19) and (20). Thus, we can simplify Problem 2 as follows.

*Problem 4:* Under the assumptions of Problem 1, find optimal communication strategy $(c^{*T}, g_1^{*T}, g_2^{*T})$ with encoder of the form (19), inner decoder of the form (20) and outer decoder of the form (3).

### C. Compressing $Z^{t-1}$ to a sufficient statistic

To find a sufficient statistic for $Z^{t-1}$, we follow the three stage approach that we followed to find a sufficient statistic for $Y^{t-1}$. These stages are

1. Formulate a stochastic control problem from the point of a coordinator that observes $Z^{t-1}$.
2. Show that the coordinator's problem is equivalent to the original problem. Specifically, any strategy for the coordinator's problem can be implemented in the original problem and vice versa.
3. Identify a controlled Markov process that is observed at the controller and use that to identify a sufficient statistic for $Z^{t-1}$.

Below we elaborate on each of these stages.

*Stage 1*

This stage is similar to stage 1 for compressing $(Y^{t-1}, Z^{t-1})$. We consider a modified problem with a coordinator that observes $Z^{t-1}$. This information $Z^{t-1}$ is the common shared information between the encoder and the two decoders. Based on this information, the coordinator decides action $\hat{V}_{t-1}$ and the partial functions

$$\hat{c}_t : \mathcal{U} \times \mathcal{V} \times \Delta(\mathcal{U} \times \mathcal{V}) \mapsto \mathcal{X},$$

$$\hat{g}_{1,t} : \mathcal{Y} \times \Delta(\mathcal{U} \times \mathcal{V}) \mapsto \hat{\mathcal{U}}.$$

These functions map the private information of the encoder and the inner decoder to their decisions. The coordinator then informs the encoder and the decoders of $\hat{V}_{t-1}$, $\hat{c}_t$, and $\hat{g}_{1,t}$. The outer decoder uses $\hat{V}_{t-1}$ as its estimate; the encoder and the inner decoder use their respective partial functions to choose an action as follows

$$X_t = \hat{c}_t(U_t, V_t, \xi_{t-1}), \tag{21}$$

$$\hat{U}_t = \hat{g}_{1,t}(Y_t, \xi_{t-1}). \tag{22}$$

The source and the channel dynamics are the same as in the original problem. At the next time step, the coordinator observes $Z_t$ and selects action $\hat{V}_t$ and partial functions $(\hat{c}_{t+1}, \hat{g}_{1,t+1})$. The system proceeds sequentially in this manner until horizon $T$. The block diagram of the system is shown in Figure 4.

In the above formulation, there is only one decision maker: the coordinator. The encoder and the decoders simply carry out the computations prescribed in (21) and (22). The coordinator's decision rule $\hat{\phi}_t$ maps its information at time $t$ to its decision, that is,

$$(\hat{V}_{t-1}, \hat{c}_t, \hat{g}_{1,t}) = \hat{\phi}_t(Z^{t-1}, \hat{V}^{t-2}, \hat{c}^{t-1}, \hat{g}_1^{t-1}). \tag{23}$$

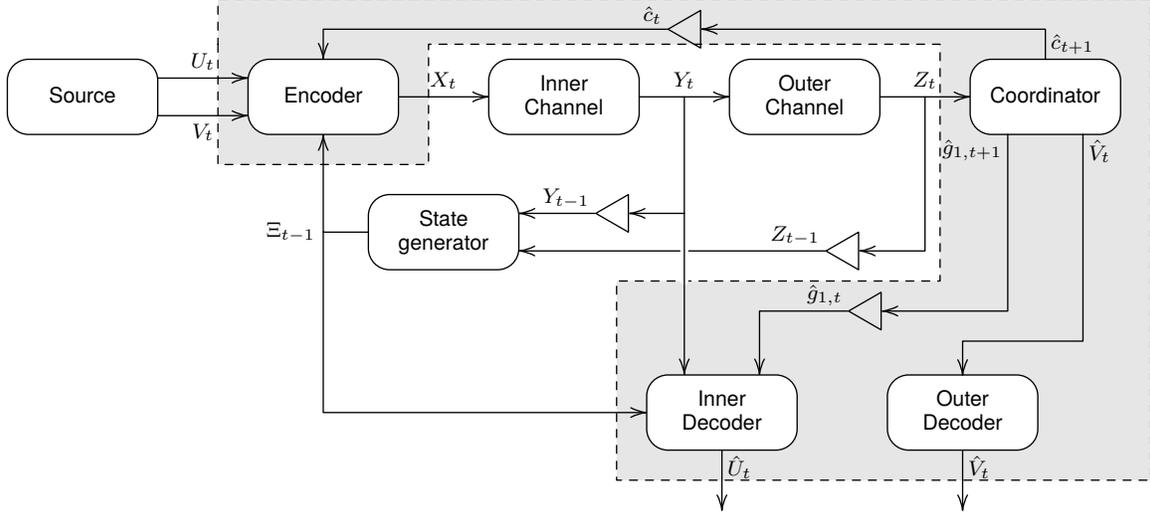

Fig. 4. Another alternate formulation of the broadcast system with feedback

The choice of $\hat{\phi}^T$ is called a coordination strategy. The expected total distortion of a strategy $\hat{\phi}^T$ is given by

$$\hat{J}(\hat{\phi}^T) = \mathbb{E}^{\hat{\phi}^T}\left\{\sum_{t=1}^{T}\left[\rho_{1,t}(U_t, \hat{U}_t) + \rho_{2,t}(V_t, \hat{V}_t)\right]\right\}. \quad (24)$$

We are interested in the optimal design of the above system, which can be set up as the following optimization problem.

*Problem 5:* Under the assumptions of Problem 1, find a strategy $\hat{\phi}^{*T}$ of the form (23) that minimizes the expected total distortion given by (24).

*Stage 2*

By an argument similar to the argument presented in Stage 2 for compressing $(Y^{t-1}, Z^{t-1})$, Problem 5 is equivalent to Problem 4. Specifically, any communication strategy $(c^T, g_1^T, g_2^T)$ for Problem 4 can be implemented by the coordinator in Problem 5 and vice versa. Thus, we can focus on deriving structural results for Problem 5.

*Stage 3*

We first define the following.

*Definition 2:* For any choice of $\hat{c}^T$, define

$$\Pi_t(Z^t; \hat{c}^t) := \Pr^{\hat{c}^t}\left(U_t, V_t, \Xi_t \mid Z^t\right). \quad (25)$$

$\Pi_t$ has the same interpretation as $\Xi_t$ defined in Definition 1. For any choice of $\hat{c}^T$, $Z^t$ is a random vector (measurable on the probability space on which the source outputs and the channel noise are defined). Given a realization $z^t$ of $Z^t$, the realization $\pi_t$ of $\Pi_t$ is a conditional probability on $(U_t, V_t)$ given $z^t$. On the other hand, when $Z^t$ is random, $\Pi_t$ is a random variable taking values in $\Delta(\mathcal{U} \times \mathcal{V})$. Furthermore, $\Pi_t$ is related to $\Pi_{t-1}$ as follows.

*Proposition 4:* Fix arbitrary partial encoding functions $\hat{c}^t$. Then the update of $\Pi_t$ is given by

$$\pi_t(z^t; \hat{c}^t) = f_2(\pi_{t-1}(z^{t-1}; \hat{c}^{t-1}); z_t, \hat{c}_t) \quad (26)$$

where

$$f_2(\pi, z, \hat{c})(u, v, \xi) = \sum_{(u', v') \in \mathcal{U} \times \mathcal{V}} P_{UV}(u, v | u', v')$$

$$\times \frac{\int_{\mathcal{U} \times \mathcal{V}} \sum_{y' \in \hat{\mathcal{Y}}(\xi, \xi', z, \hat{c})} R(u', v', y', z, \xi)}{\int_{\mathcal{U} \times \mathcal{V}} \sum_{y' \in \hat{\mathcal{Y}}(\xi, \xi', z, \hat{c})} \sum_{z' \in \mathcal{Z}} R(u', v', y', z', \xi)}.$$

with

$$R(u, v, y, z, \xi) = Q_{Z|Y}(z|y)Q_{Y|X}(y|\hat{c}(u, v))\xi(u, v)$$

and $\hat{\mathcal{Y}}(\xi, \xi', z, \hat{c}) := \{y \in \mathcal{Y} : \xi = f_2(\xi', y, z, \hat{c})\}$.

*Proof.* This is a direct consequence of Definition 2 and Bayes's rule. □

$\Pi_{t-1}$ is a sufficient statistic for $Z^{t-1}$ in Problem 5. In particular, we have the following result.

*Proposition 5:* In Problem 5, without loss of optimality we can restrict attention to a coordination strategy of the form

$$\hat{\phi}_t : \Delta(\mathcal{U} \times \mathcal{V} \times \Delta(\mathcal{U} \times \mathcal{V}))$$
$$\mapsto (\hat{\mathcal{V}}, (\mathcal{U} \times \mathcal{V} \times \Delta(\mathcal{U} \times \mathcal{V}) \mapsto \mathcal{X}), (\mathcal{Y} \times \Delta(\mathcal{U} \times \mathcal{V}) \mapsto \hat{\mathcal{U}}))$$

with

$$(\hat{V}_{t-1}, \hat{c}_t, \hat{g}_{1,t}) = \hat{\phi}_t(\Pi_{t-1}). \quad (27)$$

Since Problems 4 and 5 are equivalent, the above implies that in Problem 4, without loss of optimality, we can restrict attention to communication strategies of the form

$$c_t : \mathcal{U} \times \mathcal{V} \times \Delta(\mathcal{U} \times \mathcal{V}) \times \Delta(\Delta(\mathcal{U} \times \mathcal{V})) \mapsto \mathcal{X},$$
$$g_{1,t} : \mathcal{Y} \times \Delta(\mathcal{U} \times \mathcal{V}) \times \Delta(\Delta(\mathcal{U} \times \mathcal{V})) \mapsto \hat{\mathcal{U}},$$
$$g_{2,t} : \Delta(\Delta(\mathcal{U} \times \mathcal{V})) \mapsto \hat{\mathcal{V}}.$$

with

$$X_t = c_t(U_t, V_t, \Xi_{t-1}, \Pi_{t-1}), \quad (28)$$
$$\hat{U}_t = g_{1,t}(Y_t, \Xi_{t-1}, \Pi_{t-1}), \quad (29)$$
$$\hat{V}_t = g_{2,t}(\Pi_t). \quad (30)$$

*Proof.* It can be verified that

$$\Pr\left(\Pi_t \mid \Pi^{t-1}; \hat{c}^t, \hat{g}_1^t, \hat{g}_2^t\right) = \Pr\left(\Pi_t \mid \Pi_{t-1}; \hat{c}_t\right).$$

Furthermore,

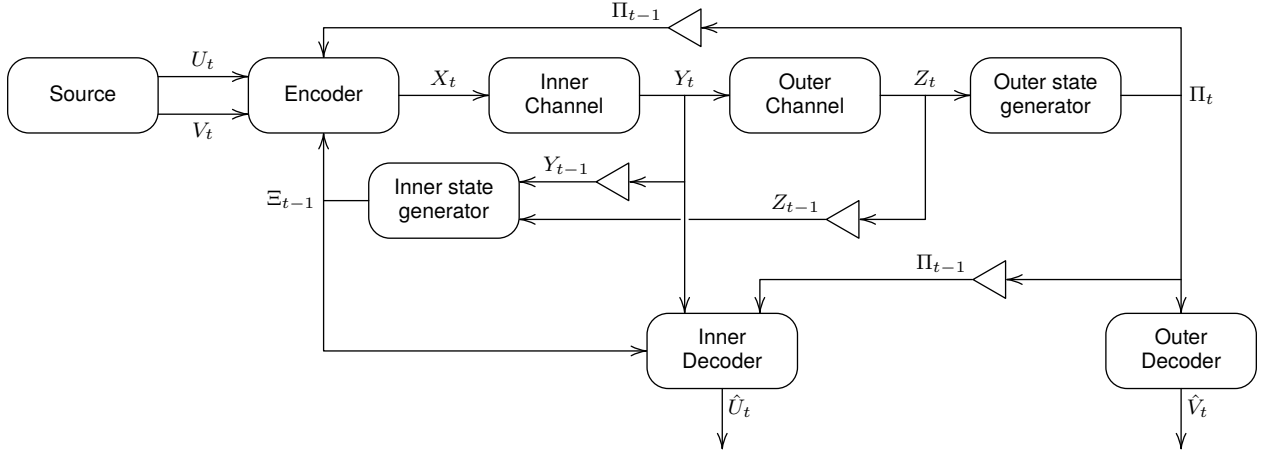

Fig. 5. The broadcast system with simplified encoder and inner decoder and outer decoder.

$$\mathbb{E}\left\{\rho_{1,t}(U_t,\hat{U}_t)+\rho_{2,t}(V_t,\hat{V}_t)\,\Big|\,\Pi^{t-1};\hat{c}^t,\hat{g}_1^t,\hat{g}_2^t\right\}$$
$$=\mathbb{E}\left\{\rho_{1,t}(U_t,\hat{U}_t)+\rho_{2,t}(V_t,\hat{V}_t)\,\Big|\,\Pi_{t-1};\hat{c}_t,\hat{g}_{1,t},\hat{g}_{2,t}\right\}$$

Thus, the expected instantaneous cost only depends on $(\Pi_{t-1}, \hat{c}_t, \hat{g}_{1,t}, \hat{g}_{2,t})$. Moreover, the state $\Pi_{t-1}$ is perfectly observed at the coordinator. Hence, the results of Markov decision theory [6] imply that restricting attention to coordinator strategies of the form (27) does not incur a loss of optimality. □

From now on, we will assume that the encoder and the decoders are of the form (28), (29), and (30). Thus, the broadcast system can be viewed as shown in Figure 5. The system has a *outer state-generator*, which carries out the computations prescribed in (26). At time $t$, the state generator computes $\Pi_t$. $\Pi_t$ is immediately communicated to the outer decoder, and it is communicated with a unit delay to the encoder and the inner decoder. The encoder and the decoders use $\Pi_t$ along with their private information to implement communication strategy of the form (28), (29) and (30).

### D. A smaller sufficient statistic for the decoders

The results of Proposition 5 show that we can restrict attention to encoders and decoders that have a time-invariant domain. The decoders can be further simplified by exploiting the fact that the decoding is a filtration, i.e., the decoder's decision do not affect the future evolution of the system. For that matter, we define the following.

*Definition 3:* For any choice of $c^T$, define

$$\theta_{1,t}(Y^t,Z^{t-1};c^t)=\mathrm{Pr}^{c^t}\!\left(U_t\,\Big|\,Y^t,Z^{t-1}\right), \quad (31)$$
$$\theta_{2,t}(Z^t;c^t)=\mathrm{Pr}^{c^t}\!\left(V_t\,\Big|\,Z^t\right). \quad (32)$$

For any choice of $c^t$, $(Y^t,Z^t)$ are random variables (measurable on the probability space on which the source outputs and the channel noise are defined). Given a realization of $(y^t,z^{t-1})$, the realization $\theta_{1,t}$ of $\Theta_{1,t}$ is conditional probability on $U_t$ given $(y^t,z^{t-1})$. Similarly, given a realization of $z^t$, the realization $\theta_{2,t}$ of $\Theta_{2,t}$ is a conditional probability on $V_t$ given $z^t$. On the other hand, when $(Y^t,Z^t)$ are random variables, $\Theta_{1,t}$ and $\Theta_{2,t}$ are random variables taking value in $\Delta(\mathcal{U})$ and $\Delta(\mathcal{V})$. Moreover, $\Theta_{1,t}$ and $\Theta_{2,t}$ are related to $\Xi_{t-1}$ and $\Pi_{t-1}$ as follows.

*Proposition 6:* Fix arbitrary encoding functions $c^T$ of the form (28). Let $\hat{c}_t(\cdot,\cdot)=c_t(\cdot,\cdot,\xi_{t-1},\pi_{t-1})$. Then, $\Theta_{1,t}$ and $\Theta_{2,t}$ are given by

$$\theta_{1,t}(y^t,z^{t-1};c^t)=h_1\big(\xi_{t-1}(y^{t-1},z^{t-1};\hat{c}^{t-1}),y_t,\hat{c}_t\big), \quad (33)$$
$$\theta_{2,t}(z^t;c^t)=h_2\big(\pi_t(z^t;\hat{c}^t)\big); \quad (34)$$

where

$$h_1(\xi,y,\hat{c})(u)=\sum_{v'\in\mathcal{V}}\frac{Q_{Y|X}(y|\hat{c}(u,v',\xi))\xi(u,v')}{\sum_{y'\in\mathcal{Y}}Q_{Y|X}(y'|\hat{c}(u,v'))\xi(u,v')},$$
$$h_2(\pi)(v)=\sum_{u'\in\mathcal{U}}\int_{\Delta(\mathcal{U}\times\mathcal{V})}\pi(u',v,\xi')\,d\xi'.$$

*Proof.* This is a direct consequence of the definitions of $\Theta_{1,t}$, $\Theta_{2,t}$, $\Xi_t$, $\Pi_t$, and Baye's rule. □

$\Theta_{1,t}$ and $\Theta_{2,t}$ are sufficient statistics for the decoders. Specifically,

*Proposition 7:* Without loss of optimality, we can restrict attention to decoders of the form

$$\hat{U}_t=\tau_{1,t}(\Theta_{1,t}) \quad (35)$$
$$\hat{V}_t=\tau_{2,t}(\Theta_{2,t}) \quad (36)$$

where

$$\tau_{1,t}(\theta_1)=\arg\min_{\hat{u}\in\hat{\mathcal{U}}}\sum_{u\in\mathcal{U}}\rho_{1,t}(u,\hat{u})\theta_1(u),$$

and

$$\tau_{2,t}(\theta_2)=\arg\min_{\hat{v}\in\hat{\mathcal{V}}}\sum_{v\in\mathcal{V}}\rho_{2,t}(v,\hat{v})\theta_2(v).$$

*Proof.* For any arbitrary but fixed choice of the encoding and outer decoding rule, the choice of decoding rules is a filtration, i.e., the choice of decoded symbols does not affect the future evolution of the system. Hence, the inner and outer decoders can choose $\hat{U}_t$ and $\hat{V}_t$ to minimize $\mathbb{E}\left\{\rho_{1,t}(U_t,\hat{U}_t)\,\Big|\,Y^t,Z^{t-1}\right\}$ and $\mathbb{E}\left\{\rho_{2,t}(V_t,\hat{V}_t)\,\Big|\,Z^t\right\}$, respectively. Consequently, optimal decoders can be of the form (35) and (36). □

From now on, we will assume that the decoders are of the form (35) and (36). Thus, the broadcast system can be viewed as shown in Figure 6. There are two modifications. First, the system has two extra components, the inner and outer *state-compressors*. Second, the outer state compressor communicates communicates $\hat{c}_t$ to the encoder and the inner decoder while it communicates $\Pi_t$ to the outer state-compressor. The state-compressors carry out the computations of (33) and (34). At time $t$, they compute $\Theta_{i,t}$, $i=1,2$, and communicate these to their corresponding decoders. The decoders

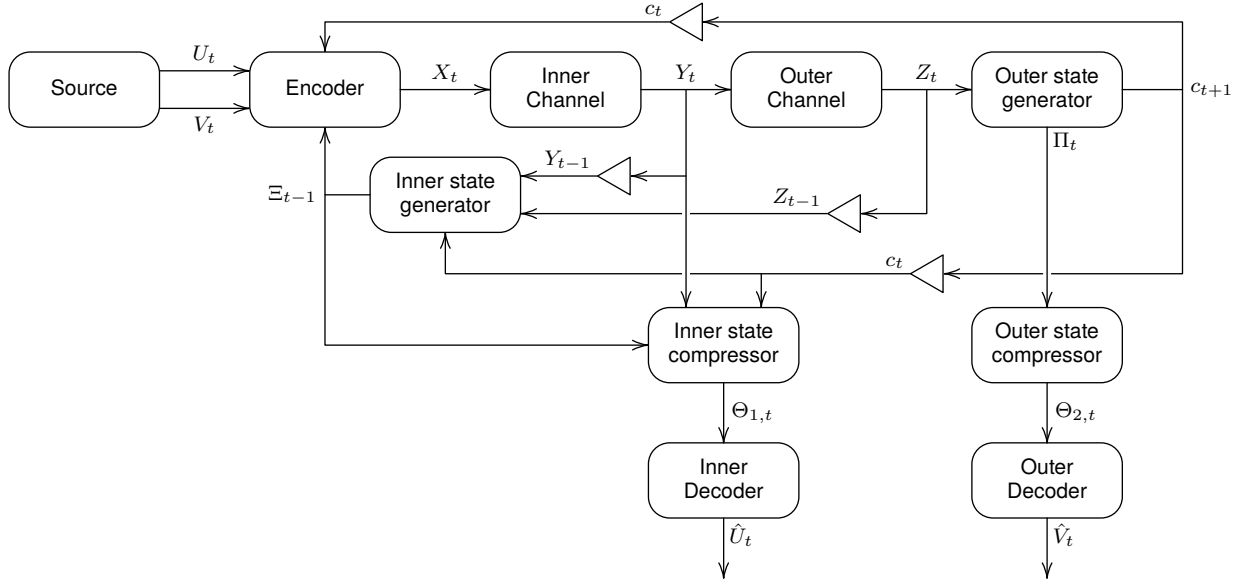

Fig. 6. The broadcast system with simplified inner decoder and outer decoder.

use $\Theta_{i,t}$ and generate $U_t$ and $V_t$ according to $\tau_{i,t}$.

### E. Combined structural results

We can combine the results of the previous sections to get Theorem 1. We restate a more detailed version of that theorem below.

*Theorem 2:* Without loss of optimality, we can restrict attention to communication strategies where the encoding rule is of the form

$$c_t : \mathcal{U} \times \mathcal{V} \times \Delta(\mathcal{U} \times \mathcal{V}) \times \Delta\big(\Delta(\mathcal{U} \times \mathcal{V})\big) \mapsto \mathcal{X}, \qquad (37)$$

and the decoding functions are of the form

$$g_{1,t} : \Delta(\mathcal{U}) \mapsto \hat{\mathcal{U}}, \qquad (38)$$
$$g_{2,t} : \Delta(\mathcal{V}) \mapsto \hat{\mathcal{V}}. \qquad (39)$$

The encoder and the decoders operate as follows:

$$X_t = c_t(U_t, V_t, \Xi_{t-1}, \Pi_{t-1}) = \hat{c}_t(\Pi_{t-1})(U_t, V_t, \Xi_{t-1}), \qquad (40)$$

and

$$\hat{\mathcal{U}}_t = \tau_{1,t}(h_1(\Xi_{t-1}, Y_t, \hat{c}_t)), \qquad (41)$$
$$\hat{\mathcal{V}}_t = \tau_{2,t}(h_2(\Pi_t)); \qquad (42)$$

where $\tau_{1,t}$ and $\tau_{2,t}$ are defined in Proposition 7 and $h_1$ and $h_2$ are defined in Proposition 6.

## III. Conclusion

We presented structural properties of optimal encoders and decoders for sequential transmission over degraded broadcast channel with nested feedback. Our technical approach is based on ideas from decentralized team theory. We obtain the structural results by a sequence of steps; each step compresses an increasing sequence of observations into a sufficient statistic taking values in a fixed space. To identify these sufficient statistics, we identify coordinators for two or more agents that observes the common information of these agents. We believe that this idea formulating equivalent problem from the point of view of a coordinator observing common information is also useful in other multi-terminal communication problems.

## Acknowledgment

The author is grateful to Achilleas Anastasopoulos for bringing this problem to his attention and for insightful discussions.